\documentclass[11pt]{article}
\pdfoutput=1

\usepackage{amsmath,amssymb,hyperref}
\usepackage{fullpage}

\begin{document}

\title{A counterexample to the Nelson-Seiberg theorem}
\author{Zheng Sun\textsuperscript{*}, Zipeng Tan\textsuperscript{\dag}, Lu Yang\textsuperscript{\ddag}\\
        \normalsize\textit{College of Physics, Sichuan University, 29 Wangjiang Road, Chengdu 610064, P.~R.~China}\\
        \normalsize\textit{E-mail:}
        \textsuperscript{*}\texttt{sun\_ctp@scu.edu.cn,}
        \textsuperscript{\dag}\texttt{tzpcyc@126.com,}
        \textsuperscript{\ddag}\texttt{1546265328@qq.com}
       }
\date{}
\maketitle

\begin{abstract}
We present a counterexample to the Nelson-Seiberg theorem and its extensions.  The model has $4$ chiral fields, including one R-charge $2$ field and no R-charge $0$ filed.  Giving generic values of coefficients in the renormalizable superpotential, there is a supersymmetric vacuum with one complex dimensional degeneracy.  The superpotential equals zero and the R-symmetry is broken everywhere on the degenerated vacuum.  The existence of such a vacuum disagrees with both the original Nelson-Seiberg theorem and its extensions, and can be viewed as the consequence of a non-generic R-charge assignment.  Such counterexamples may introduce error to the field counting method for surveying the string landscape, and are worth further investigations.
\end{abstract}

\section{Introduction}

In 4-dimentinal $N = 1$ supersymmetry (SUSY) theories~\cite{Martin:1997ns, Intriligator:2007cp}, the relation between SUSY breaking and R-symmetries in Wess-Zumino models are described by the Nelson-Seiberg theorem and its extensions~\cite{Nelson:1993nf, Sun:2011fq, Kang:2012fn, Li:2020wdk}.  The original Nelson-Seiberg theorem~\cite{Nelson:1993nf} states that for SUSY breaking at a stable vacuum in a generic model, a necessary condition is to have an R-symmetric superpotentical, and a sufficient condition is to have the R-symmtery spontaneously broken at the vacuum.  The statement can be extended to metastable SUSY breaking models with approximate R-symmetries~\cite{Intriligator:2006dd, Intriligator:2007py, Abe:2007ax}.  A revised theorem~\cite{Kang:2012fn} claims that with a polynomial superpotential, the neccessary and sufficient condition for SUSY breaking is to have an R-symmetric superpotentical and more R-charge $2$ fields than R-charge $0$ fields.  The revised theorem can be generalized to include models with non-polynomial superpotentials \cite{Li:2020wdk}, where the condition for SUSY breaking is the same as before for superpotentials smooth at the origin of the field space, and a singularity at the origin implies both R-symmetry breaking and supersymmetry breaking.  Half of the revised theorem has even stronger claims~\cite{Sun:2011fq}:  If we have an R-symmetric superpotentical and less or equal R-charge $2$ fields than R-charge $0$ fields, the SUSY vacua claimed by the revised theorem have an additional property that superpotentials equal zero at the vacuum, and this part of claim is also true for non-$\mathbb{Z}_2$ discrete R-symmetries or non-Abelian discrete R-symmetries~\cite{Chen:2013dpa, Brister:2020xxx}.

The Nelson-Seiberg theorem and its extensions provides various tools for new physics model building.  In SUSY phenomenology beyond the Standard Model, the SUSY breaking effect is mediated to the SUSY Standard Model sector through a messenger sector.  So constructing the SUSY breaking sector with R-symmetries makes the first step towards a full model.  In string phenomenology, flux compactification of type IIB string theory~\cite{Grana:2005jc, Douglas:2006es, Blumenhagen:2006ci, Denef:2007pq} gives low energy effective theories formulated as a supergravity (SUGRA) version of Wess-Zumino models, and discrete R-symmetries come from geometrical symmetries of the Calabi-Yau manifold used for compactification.  The previous condition for SUSY vacua with zero superpotentials gives again SUSY vacua with zero vacuum energy in SUGRA\@.  SUSY breaking and the vacuum energy are then non-perturbatively generated at lower scales, notably through the racetrack mechanism~\cite{Krasnikov:1987jj}.  Such vacua contribute to the third branch of the string landscape which prefers low scale SUSY breaking~\cite{Dine:2004is, Dine:2005yq, Dine:2005gz}.  And the de Sitter swampland conjecture~\cite{Obied:2018sgi, Garg:2018reu, Ooguri:2018wrx, Palti:2019pca} may hopefully be evaded by these vacua.  Following the revised Nelson-Seiberg theorem, whether a SUSY vacuum exists or not can be generically determined by counting R-charge $2$ and R-charge $0$ fields, and the explicit vacuum solution is not needed.  Such field counting method makes it possible to do a fast survey of the string landscape.

Both the original Nelson-Seiberg theorem and its extensions require genericness assumptions.  The usual concept of genericness means that parameters take generic values.  It is related to naturalness, fine-tuning or hierarchy problems~\cite{tHooft:1979rat, Giudice:2008bi, Feng:2013pwa, Dine:2015xga}.  This type of non-generic models only compose a null set in the parameter space, and can be neglected in both phenomenology or string phenomenology studies.  Another lesser-known concept of genericness is about R-charges.  With some special R-charge assignment which determines a renormalizable superpotential respecting the R-symmetry, the explicit vacuum solution disagrees with what the previous theorems predict.  These models still have generic parameters, and can be viewed as having a non-generic R-charge assignment.  This work is to present the first known model of this type as a constructive proof.  If such non-genericness occurs often, it may introduce non-neglectable error to the field counting method from the revised Nelson-Seiberg theorem.

The rest of this paper is organized as follows.  Section 2 presents the model and its vacuum structure, showing that it is a counterexample to both the original Nelson-Seiberg theorem and the revised theorem, with non-generic R-charges.  Section 3 discusses properties of the SUSY vacuum and implications for both phenomenology and string phenomenology studies.

\section{The counterexample}

The model presented here has four chiral fields $\{z_1, z_2, z_3, z_4\}$ with the R-charge assignment
\begin{equation} \label{eq:2-01}
R(z_1) = 2, \quad
R(z_2) = - 2, \quad
R(z_3) = 6, \quad
R(z_4) = - 6.
\end{equation}
In the following context, the same set of symbols are used for chiral fields and their scalar components which determine vacuum solutions.  A renormalizable R-symmetric superpotential has the form
\begin{equation} \label{eq:2-02}
W = a z_1 + b z_1^2 z_2 + c z_2^2 z_3 + d z_1 z_3 z_4,
\end{equation}
where the coefficients $a$, $b$, $c$ and $d$ take generic complex values.  All R-charges of fields are uniquely fixed by requiring $W$ to have R-charge $2$, and all R-charge $2$ monomial terms up to cubic are included in $W$.  Following the field counting method used in~\cite{Sun:2011fq, Kang:2012fn}, the number of R-charge $2$ fields $N_X = 1$ is greater than the number of R-charge $0$ fields $N_Y = 0$.  The revised Nelson-Seiberg theorem~\cite{Kang:2012fn} claims that there is no SUSY vacuum.  But solving the SUSY equations
\begin{align}
\partial_1 W &= a + 2 b z_1 z_2 + d z_3 z_4 = 0,\\
\partial_2 W &= b z_1^2 + 2 c z_2 z_3 = 0,\\
\partial_3 W &= c z_2^2 + d z_1 z_4 = 0,\\
\partial_4 W &= d z_1 z_3 = 0
\end{align}
gives a SUSY vacuum at
\begin{equation} \label{eq:2-03}
z_1 = z_2 = 0, \quad z_3 z_4 = - \frac{a}{d},
\end{equation}
which also satisfies an additional equation $W = 0$.  The vacuum has one complex dimensional degeneracy on the $z_3$-$z_4$ space for non-zero $a$ and $d$.  Since $z_3$ and $z_4$ has non-zero R-charges, the R-symmetry is spontaneous broken everywhere on the degenerated vacuum, which should lead to SUSY breaking according to the original Nelson-Seiberg theorem~\cite{Nelson:1993nf}.  The existence of the SUSY vacuum means that this model is a counterexample to both the original Nelson-Seiberg theorem and the revised one.  The SUSY solution is a result of the particular form of the renormalizable superpotential~\eqref{eq:2-02}, which is constructed from fields with the particular R-charges~\eqref{eq:2-01}.  Thus the counterexample presented here can be viewed as having a non-generic R-charge assignment.

The full vacuum structure of the model may be obtained from the scalar potential
\begin{equation}
V = (\partial_i W)^* \partial_i W,
\end{equation}
where a minimal K\"ahler potential is assumed, and the Einstein summation convention for field indices is adopted.  Stationary points are found by solving the zero points of the first derivatives of $V$:
\begin{equation}
\partial_i V = (\partial_j W)^* \partial_i \partial_j W = 0.
\end{equation}
Whether a stationary point is a minimum, maximum or saddle point can be determined by checking the eigenvalues of the second derivative matrix of $V$:
\begin{equation}
\partial^2 V =
\begin{pmatrix}
\partial_{\bar i} \partial_j V & \partial_{\bar i} \partial_{\bar j} V\\
\partial_i \partial_j V        & \partial_i \partial_{\bar j} V
\end{pmatrix}
=
\begin{pmatrix}
(\partial_i \partial_k W)^* \partial_j \partial_k W & (\partial_i \partial_j \partial_k W)^* \partial_k W\\
(\partial_k W)^* \partial_i \partial_j \partial_k W & (\partial_j \partial_k W)^* \partial_i \partial_k W
\end{pmatrix}.
\end{equation}
There is a saddle point at the origin of the field space
\begin{equation}
z_1 = z_2 = z_3 = z_4 = 0.
\end{equation}
Using the complexified symmetry technique developed in~\cite{Ferretti:2007ec, Ferretti:2007rq, Azeyanagi:2012pc, Sun:2018hnk}, SUSY runaways are found along the direction
\begin{gather}
z_1 = - \frac{c}{d} u v, \quad
z_2 = \frac{v}{u}, \quad
z_3 = (\frac{2 b c}{d^2} v^2 - \frac{a}{d}) \frac{u^3}{v}, \quad
z_4 = \frac{v}{u^3},\\
v = \sqrt \frac{a d (10 + 2 \lvert u \rvert^4)}{b c (25 + 4 \lvert u \rvert^4)}, \quad
u \to 0.
\end{gather}
Further analytical search for stationary point easily exhausts our available computational resource.  Numerical calculation with typical coefficient values shows that several degenerated saddle points exist in addition to the one at the origin.  But no local minimum other than the SUSY one has been found so far after extensive search.  The SUSY vacuum~\eqref{eq:2-03} is most likely the only metastable (and stable) minimum in our model.

\section{Discussions}

Considering non-renormalizable superpotentials, there are simpler models than ours.  For example, the quartic superpotential
\begin{equation}
W = a z_1 + b z_1^3 z_3 + c z_1 z_2 z_3
\end{equation}
with the the R-charge assignment
\begin{equation}
R(z_1) = 2, \quad
R(z_2) = 4, \quad
R(z_3) = - 4,
\end{equation}
gives a similar vacuum structure with a SUSY vacuum at
\begin{equation}
z_1 = 0, \quad z_2 z_3 = - \frac{a}{c}.
\end{equation}
This model is also a counterexample to both the original Nelson-Seiberg theorem and the revised one.  The quartic term has to be included to uniquely fix all R-charges.  If the superpotential is restricted to cubic to be renormalizable, at least $4$ fields are needed to construct a counterexample and our model is the simplest one.

It is worth to note that a different type of R-symmetry breaking SUSY vacua already exist in some previously known examples with non-generic superpotentials~\cite{Ellis:1982vi}, or generic models with non-renormalizable superpotentials~\cite{Ray:2007wq}.  Their degenerated SUSY vacua contain the origin of the field space which preserves the R-symmetry.  Since the Nelson-Seiberg theorem gives no information on the existence of a vacuum at the origin, those models do not really contradict with the theorem.  Moreover, the models in~\cite{Ray:2007wq} have less or equal R-charge $2$ fields than R-charge $0$ fields.  So the field counting method gives correct prediction on the SUSY vacua.  On the contrary, our model has R-symmetry breaking everywhere on the degenerated SUSY vacuum~\eqref{eq:2-03}, and more R-charge $2$ fields than R-charge $0$ fields.  Thus it is the first known counterexample to both the original Nelson-Seiberg theorem and the revised one, with a non-generic R-charge assignment and a generic renormalizable superpotential respecting the R-symmetry.

Because of the R-symmetry breaking feature of the vacuum, our model can serve as a tree-level R-symmetry breaking sector separate from a SUSY breaking sector.  It is distinct from previous tree-level R-symmetry breaking models where both SUSY breaking and R-symmetry breaking happen in the same sector~\cite{Shih:2007av, Carpenter:2008wi, Sun:2008va, Komargodski:2009jf, Curtin:2012yu}.  The separation of SUSY breaking and R-symmetry breaking may gives the possibility to generate the gaugino mass from tree-level R-symmetry breaking in gauge mediation models~\cite{Liu:2014ida}.

The superpotential~\eqref{eq:2-02} vanishes at the SUSY vacuum~\eqref{eq:2-03}.  Similarly to the case with less or equal R-charge $2$ fields than R-charge $0$ fields~\cite{Sun:2011fq}, the SUGRA version of the model gives again a SUSY vacuum with zero vacuum energy, and contributes to the the third branch of the string landscape.  Since the model has more R-charge $2$ fields than R-charge $0$ fields. the revised Nelson-Seiberg theorem does not predict a SUSY vacuum.  But when realized in flux compactification of type IIB string theory~\cite{Grana:2005jc, Douglas:2006es, Blumenhagen:2006ci, Denef:2007pq}, the R-symmetry breaking feature of the vacuum means that the expectation values of moduli has sent the Calabi-Yau manifold away from the R-symmetric point of its moduli space.  It is then unnatural to turn on only R-symmetric fluxes and obtain an R-symmetric effective superpotential.  Our model does not affect the accuracy of the field counting method if we only consider R-symmetric SUSY vacua in the third branch, or string vacua with enhanced symmetries~\cite{DeWolfe:2004ns, DeWolfe:2005gy, Palti:2007pm, Kanno:2017nub, Palti:2020qlc}.  A recent exploration of new counterexamples~\cite{Amariti:2020lvx} shows that such R-symmetry breaking feature of SUSY vacua is common in all currently known constructions.  Whether other types of counterexamples exist is worth further investigations for an accurate survey of the string landscape.

\section*{Acknowledgement}

The authors thank Shihao Kou and Zhengyi Li for helpful discussions. This work is supported by the National Natural Science Foundation of China under the grant number 11305110.


\begin{thebibliography}{9}

\bibitem{Martin:1997ns}
S.~P.~Martin,
``A Supersymmetry primer,''
Adv. Ser. Direct. High Energy Phys. \textbf{21} (2010), 1-153
doi:10.1142/9789812839657\_0001
[arXiv:hep-ph/9709356 [hep-ph]].

\bibitem{Intriligator:2007cp}
K.~A.~Intriligator and N.~Seiberg,
``Lectures on Supersymmetry Breaking,''
Class. Quant. Grav. \textbf{24} (2007), S741-S772
doi:10.1088/0264-9381/24/21/S02
[arXiv:hep-ph/0702069 [hep-ph]].

\bibitem{Nelson:1993nf}
A.~E.~Nelson and N.~Seiberg,
``R symmetry breaking versus supersymmetry breaking,''
Nucl. Phys. B \textbf{416} (1994), 46-62
doi:10.1016/0550-3213(94)90577-0
[arXiv:hep-ph/9309299 [hep-ph]].

\bibitem{Sun:2011fq}
Z.~Sun,
``Low energy supersymmetry from R-symmetries,''
Phys. Lett. B \textbf{712} (2012), 442-444
doi:10.1016/j.physletb.2012.05.013
[arXiv:1109.6421 [hep-th]].

\bibitem{Kang:2012fn}
Z.~Kang, T.~Li and Z.~Sun,
``The Nelson-Seiberg theorem revised,''
JHEP \textbf{12} (2013), 093
doi:10.1007/JHEP12(2013)093
[arXiv:1209.1059 [hep-th]].

\bibitem{Li:2020wdk}
Z.~Li and Z.~Sun,
``The Nelson-Seiberg theorem generalized with nonpolynomial superpotentials,''
Adv. High Energy Phys. \textbf{2020} (2020), 3701943
doi:10.1155/2020/3701943
[arXiv:2006.00538 [hep-th]].

\bibitem{Intriligator:2006dd}
K.~A.~Intriligator, N.~Seiberg and D.~Shih,
``Dynamical SUSY breaking in meta-stable vacua,''
JHEP \textbf{04} (2006), 021
doi:10.1088/1126-6708/2006/04/021
[arXiv:hep-th/0602239 [hep-th]].

\bibitem{Intriligator:2007py}
K.~A.~Intriligator, N.~Seiberg and D.~Shih,
``Supersymmetry breaking, R-symmetry breaking and metastable vacua,''
JHEP \textbf{07} (2007), 017
doi:10.1088/1126-6708/2007/07/017
[arXiv:hep-th/0703281 [hep-th]].

\bibitem{Abe:2007ax}
H.~Abe, T.~Kobayashi and Y.~Omura,
``R-symmetry, supersymmetry breaking and metastable vacua in global and local supersymmetric theories,''
JHEP \textbf{11} (2007), 044
doi:10.1088/1126-6708/2007/11/044
[arXiv:0708.3148 [hep-th]].

\bibitem{Chen:2013dpa}
M.~C.~Chen, M.~Ratz and A.~Trautner,
JHEP \textbf{09} (2013), 096
doi:10.1007/JHEP09(2013)096
[arXiv:1306.5112 [hep-ph]].

\bibitem{Brister:2020xxx}
J.~Brister, Z.~Li and Z.~Sun,
to appear.

\bibitem{Grana:2005jc}
M.~Grana,
``Flux compactifications in string theory: A Comprehensive review,''
Phys. Rept. \textbf{423} (2006), 91-158
doi:10.1016/j.physrep.2005.10.008
[arXiv:hep-th/0509003 [hep-th]].

\bibitem{Douglas:2006es}
M.~R.~Douglas and S.~Kachru,
``Flux compactification,''
Rev. Mod. Phys. \textbf{79} (2007), 733-796
doi:10.1103/RevModPhys.79.733
[arXiv:hep-th/0610102 [hep-th]].

\bibitem{Blumenhagen:2006ci}
R.~Blumenhagen, B.~Kors, D.~Lust and S.~Stieberger,
``Four-dimensional String Compactifications with D-Branes, Orientifolds and Fluxes,''
Phys. Rept. \textbf{445} (2007), 1-193
doi:10.1016/j.physrep.2007.04.003
[arXiv:hep-th/0610327 [hep-th]].

\bibitem{Denef:2007pq}
F.~Denef, M.~R.~Douglas and S.~Kachru,
``Physics of String Flux Compactifications,''
Ann. Rev. Nucl. Part. Sci. \textbf{57} (2007), 119-144
doi:10.1146/annurev.nucl.57.090506.123042
[arXiv:hep-th/0701050 [hep-th]].

\bibitem{Krasnikov:1987jj}
N.~V.~Krasnikov,
``On Supersymmetry Breaking in Superstring Theories,''
Phys. Lett. B \textbf{193} (1987), 37-40
doi:10.1016/0370-2693(87)90452-7

\bibitem{Dine:2004is}
M.~Dine, E.~Gorbatov and S.~D.~Thomas,
``Low energy supersymmetry from the landscape,''
JHEP \textbf{08} (2008), 098
doi:10.1088/1126-6708/2008/08/098
[arXiv:hep-th/0407043 [hep-th]].

\bibitem{Dine:2005yq}
M.~Dine, D.~O'Neil and Z.~Sun,
``Branches of the landscape,''
JHEP \textbf{07} (2005), 014
doi:10.1088/1126-6708/2005/07/014
[arXiv:hep-th/0501214 [hep-th]].

\bibitem{Dine:2005gz}
M.~Dine and Z.~Sun,
``R symmetries in the landscape,''
JHEP \textbf{01} (2006), 129
doi:10.1088/1126-6708/2006/01/129
[arXiv:hep-th/0506246 [hep-th]].

\bibitem{Obied:2018sgi}
G.~Obied, H.~Ooguri, L.~Spodyneiko and C.~Vafa,
``De Sitter Space and the Swampland,''
[arXiv:1806.08362 [hep-th]].

\bibitem{Garg:2018reu}
S.~K.~Garg and C.~Krishnan,
``Bounds on Slow Roll and the de Sitter Swampland,''
JHEP \textbf{11} (2019), 075
doi:10.1007/JHEP11(2019)075
[arXiv:1807.05193 [hep-th]].

\bibitem{Ooguri:2018wrx}
H.~Ooguri, E.~Palti, G.~Shiu and C.~Vafa,
``Distance and de Sitter Conjectures on the Swampland,''
Phys. Lett. B \textbf{788} (2019), 180-184
doi:10.1016/j.physletb.2018.11.018
[arXiv:1810.05506 [hep-th]].

\bibitem{Palti:2019pca}
E.~Palti,
``The Swampland: Introduction and Review,''
Fortsch. Phys. \textbf{67} (2019) no.6, 1900037
doi:10.1002/prop.201900037
[arXiv:1903.06239 [hep-th]].

\bibitem{tHooft:1979rat}
G.~'t Hooft,
``Naturalness, chiral symmetry, and spontaneous chiral symmetry breaking,''
NATO Sci. Ser. B \textbf{59} (1980), 135-157
doi:10.1007/978-1-4684-7571-5\_9

\bibitem{Giudice:2008bi}
G.~F.~Giudice,
``Naturally Speaking: The Naturalness Criterion and Physics at the LHC,''
in ``Perspectives on LHC physics,''
edited by G.~Kane and A.~Pierce,
World Scientific (2008), p.155-178
doi:10.1142/9789812779762\_0010
[arXiv:0801.2562 [hep-ph]].

\bibitem{Feng:2013pwa}
J.~L.~Feng,
``Naturalness and the Status of Supersymmetry,''
Ann. Rev. Nucl. Part. Sci. \textbf{63} (2013), 351-382
doi:10.1146/annurev-nucl-102010-130447
[arXiv:1302.6587 [hep-ph]].

\bibitem{Dine:2015xga}
M.~Dine,
``Naturalness Under Stress,''
Ann. Rev. Nucl. Part. Sci. \textbf{65} (2015), 43-62
doi:10.1146/annurev-nucl-102014-022053
[arXiv:1501.01035 [hep-ph]].

\bibitem{Ferretti:2007ec}
L.~Ferretti,
``R-symmetry breaking, runaway directions and global symmetries in O'Raifeartaigh models,''
JHEP \textbf{12} (2007), 064
doi:10.1088/1126-6708/2007/12/064
[arXiv:0705.1959 [hep-th]].

\bibitem{Ferretti:2007rq}
L.~Ferretti,
``O'Raifeartaigh models with spontaneous R-symmetry breaking,''
AIP Conf. Proc. \textbf{957} (2007) no.1, 221-224
doi:10.1088/1742-6596/110/7/072011
[arXiv:0710.2535 [hep-th]].

\bibitem{Azeyanagi:2012pc}
T.~Azeyanagi, T.~Kobayashi, A.~Ogasahara and K.~Yoshioka,
``Runaway, D term and R-symmetry Breaking,''
Phys. Rev. D \textbf{86} (2012), 095026
doi:10.1103/PhysRevD.86.095026
[arXiv:1208.0796 [hep-ph]].

\bibitem{Sun:2018hnk}
Z.~Sun and X.~Wei,
``Runaway Directions in O’Raifeartaigh Models,''
Commun. Theor. Phys. \textbf{70} (2018) no.6, 677
doi:10.1088/0253-6102/70/6/677
[arXiv:1806.02384 [hep-th]].

\bibitem{Ellis:1982vi}
J.~R.~Ellis, L.~Smith, C.H. and G.~G.~Ross,
``Will the Universe Become Supersymmetric?,''
Phys. Lett. B \textbf{114} (1982), 227-229
doi:10.1016/0370-2693(82)90482-8

\bibitem{Ray:2007wq}
S.~Ray,
``Supersymmetric and R symmetric vacua in Wess-Zumino models,''
[arXiv:0708.2200 [hep-th]].

\bibitem{Shih:2007av}
D.~Shih,
``Spontaneous R-symmetry breaking in O'Raifeartaigh models,''
JHEP \textbf{02} (2008), 091
doi:10.1088/1126-6708/2008/02/091
[arXiv:hep-th/0703196 [hep-th]].

\bibitem{Carpenter:2008wi}
L.~M.~Carpenter, M.~Dine, G.~Festuccia and J.~D.~Mason,
``Implementing General Gauge Mediation,''
Phys. Rev. D \textbf{79} (2009), 035002
doi:10.1103/PhysRevD.79.035002
[arXiv:0805.2944 [hep-ph]].

\bibitem{Sun:2008va}
Z.~Sun,
``Tree level spontaneous R-symmetry breaking in O'Raifeartaigh models,''
JHEP \textbf{01} (2009), 002
doi:10.1088/1126-6708/2009/01/002
[arXiv:0810.0477 [hep-th]].

\bibitem{Komargodski:2009jf}
Z.~Komargodski and D.~Shih,
``Notes on SUSY and R-Symmetry Breaking in Wess-Zumino Models,''
JHEP \textbf{04} (2009), 093
doi:10.1088/1126-6708/2009/04/093
[arXiv:0902.0030 [hep-th]].

\bibitem{Curtin:2012yu}
D.~Curtin, Z.~Komargodski, D.~Shih and Y.~Tsai,
``Spontaneous R-symmetry Breaking with Multiple Pseudomoduli,''
Phys. Rev. D \textbf{85} (2012), 125031
doi:10.1103/PhysRevD.85.125031
[arXiv:1202.5331 [hep-th]].

\bibitem{Liu:2014ida}
F.~Liu, M.~Liu and Z.~Sun,
``No-go for tree-level R-symmetry breaking,''
Eur. Phys. J. C \textbf{77} (2017) no.11, 745
doi:10.1140/epjc/s10052-017-5327-2
[arXiv:1412.0183 [hep-ph]].

\bibitem{DeWolfe:2004ns}
O.~DeWolfe, A.~Giryavets, S.~Kachru and W.~Taylor,
``Enumerating flux vacua with enhanced symmetries,''
JHEP \textbf{02} (2005), 037
doi:10.1088/1126-6708/2005/02/037
[arXiv:hep-th/0411061 [hep-th]].

\bibitem{DeWolfe:2005gy}
O.~DeWolfe,
``Enhanced symmetries in multiparameter flux vacua,''
JHEP \textbf{10} (2005), 066
doi:10.1088/1126-6708/2005/10/066
[arXiv:hep-th/0506245 [hep-th]].

\bibitem{Palti:2007pm}
E.~Palti,
``Low Energy Supersymmetry from Non-Geometry,''
JHEP \textbf{10} (2007), 011
doi:10.1088/1126-6708/2007/10/011
[arXiv:0707.1595 [hep-th]].

\bibitem{Kanno:2017nub}
K.~Kanno and T.~Watari,
``Revisiting arithmetic solutions to the $W=0$ condition,''
Phys. Rev. D \textbf{96} (2017) no.10, 106001
doi:10.1103/PhysRevD.96.106001
[arXiv:1705.05110 [hep-th]].

\bibitem{Palti:2020qlc}
E.~Palti, C.~Vafa and T.~Weigand,
``Supersymmetric Protection and the Swampland,''
JHEP \textbf{06} (2020), 168
doi:10.1007/JHEP06(2020)168
[arXiv:2003.10452 [hep-th]].

\bibitem{Amariti:2020lvx}
A.~Amariti and D.~Sauro,
``On the Nelson-Seiberg theorem: generalizations and counter-examples,''
[arXiv:2005.02076 [hep-th]].
  
\end{thebibliography}
\end{document}